\documentclass[prl,aps,showpacs,epsf,superscriptaddress,twocolumn]{revtex4}

\def\be{\begin{equation}}
\def\ee{\end{equation}}

\usepackage{bm}
\usepackage{amsmath}
\usepackage[pdftex]{graphicx}
\usepackage{color}
\usepackage[usenames,dvipsnames]{xcolor}
\usepackage{mathrsfs}

\newcommand{\bea}{\begin{eqnarray}}
\newcommand{\eea}{\end{eqnarray}}
\newcommand{\bi}{\begin{itemize}}
\newcommand{\ei}{\end{itemize}}

\newcommand{\lambdavalue}{2.5}
\newcommand{\lambdaEstat}{3}
\newcommand{\lambdaEsys}{6}
\newcommand{\lambdafull}{\lambdavalue(\lambdaEstat)_{\rm stat}(\lambdaEsys)_{\rm sys}\,{\times}\,10^{-20} (\mu {\rm K})^{2} {\rm cm}^{6} {\rm s}^{-1}}
\newcommand{\lambdastat}{\lambdavalue(\lambdaEstat)_{\rm stat}\,{\times}\,10^{-20} (\mu {\rm K})^{2} {\rm cm}^{6} {\rm s}^{-1}}

\newcommand{\lambdavaluetheo}{1.52}
\newcommand{\lambdatheo}{\lambdavaluetheo\,{\times}\,10^{-20} (\mu {\rm K})^{2} {\rm cm}^{6} {\rm s}^{-1}}

\begin{document}

\title{Lifetime of the Bose Gas with Resonant Interactions}

%\author{B. Rem, A. Grier, U. Eismann, I. Ferrier-Barbut, T. Langen, F. Werner, F. Chevy, and C. Salomon }
\author{B. S. Rem}
\affiliation{Laboratoire Kastler-Brossel, \'Ecole Normale Sup\'erieure, CNRS and UPMC, 24 rue Lhomond, 75005 Paris, France}
\author{A. T. Grier}
\affiliation{Laboratoire Kastler-Brossel, \'Ecole Normale Sup\'erieure, CNRS and UPMC, 24 rue Lhomond, 75005 Paris, France}
\author{I. Ferrier-Barbut}
\affiliation{Laboratoire Kastler-Brossel, \'Ecole Normale Sup\'erieure, CNRS and UPMC, 24 rue Lhomond, 75005 Paris, France}
\author{U. Eismann}
\affiliation{Laboratoire Kastler-Brossel, \'Ecole Normale Sup\'erieure, CNRS and UPMC, 24 rue Lhomond, 75005 Paris, France}
\author{T. Langen}
\affiliation{Vienna Center for Quantum Science \& Technology, Atominstitut, TU Wien, Stadionallee 2, 1020 Wien, Austria}
\author{N. Navon}
\affiliation{Laboratoire Kastler-Brossel, \'Ecole Normale Sup\'erieure, CNRS and UPMC, 24 rue Lhomond, 75005 Paris, France}
\author{L. Khaykovich}
\affiliation{Laboratoire Kastler-Brossel, \'Ecole Normale Sup\'erieure, CNRS and UPMC, 24 rue Lhomond, 75005 Paris, France}
\affiliation{Department of Physics, Bar-Ilan University, Ramat-Gan, 52900 Israel}
\author{F.~Werner}
\affiliation{Laboratoire Kastler-Brossel, \'Ecole Normale Sup\'erieure, CNRS and UPMC, 24 rue Lhomond, 75005 Paris, France}
\author{D.~S.~Petrov}
\affiliation{Universit\'e Paris-Sud, CNRS, LPTMS, UMR8626, Orsay, F-91405, France}
\affiliation{Russian Research Center Kurchatov Institute, Kurchatov Square, 123182 Moscow, Russia}
\author{F.~Chevy}
\affiliation{Laboratoire Kastler-Brossel, \'Ecole Normale Sup\'erieure, CNRS and UPMC, 24 rue Lhomond, 75005 Paris, France}
\author{C.~Salomon}
\affiliation{Laboratoire Kastler-Brossel, \'Ecole Normale Sup\'erieure, CNRS and UPMC, 24 rue Lhomond, 75005 Paris, France}
\date{\today}

\begin{abstract}
We study the lifetime of a Bose gas at and around unitarity using a Feshbach resonance in lithium~7. At unitarity, we measure the temperature dependence of the three-body decay coefficient $L_{3}$. Our data follow a $L_3\,{=}\,\lambda_{3} / T^{2}$ law with $\lambda_{3}\,{=}\,\lambdafull$ and are in good agreement with our analytical result based on zero-range theory. Varying the scattering length $a$ at fixed temperature, we investigate the crossover between the finite-temperature unitary region and the previously studied regime where $|a|$ is smaller than the thermal wavelength. We find that $L_{3}$ is continuous across the resonance, and over the whole $a\,{<}\,0$ range our data quantitatively agree with our calculation.

\end{abstract}

 \maketitle

Recent advances in manipulating cold atomic vapors have enabled the study of Fermi gases at the unitary limit where the scattering length $a$ describing two-body interactions becomes infinite. It has been demonstrated both experimentally and theoretically that in this limit the system is characterized by a scale invariance leading to remarkably simple scaling laws \cite{zwerger2012BCSBEC}. By contrast, most experimental results on Bose-Einstein condensates were obtained in the weakly interacting regime. Recent experimental results on bosons near Feshbach resonances have revived the interest in strongly interacting bosons \cite{pollack2009extreme}: the development of experimental tools has enabled a precise test of the Lee-Huang-Yang corrections \cite{navon2011dynamics,wild2012measurements}, and several theoretical papers have studied the hypothetical unitary Bose gas at zero \cite{cowell2002cold,song2009ground,Lee2010universality,Borzov2012three} or finite \cite{HoUnitaryBose} temperature. The strongly interacting Bose gas is one of the most fundamental quantum many-body systems, yet many open questions remain. Examples include
%the continuity of the equation of state through resonance~\cite{HoUnitaryBose}, 
the prediction of weakly bound efimovian droplets~\cite{Stecher2010,Stecher2011}, the existence of both atomic and molecular superfluids~\cite{Basu2008}, and the creation of strongly correlated phases through three-body losses~\cite{Syassen2008strong}.
%Examples include the validity of universal thermodynamics~\cite{Castin2013}, the continuity of the equation of state through resonance~\cite{HoUnitaryBose}, the possible existence of both atomic and molecular superfluidity~\cite{Basu2008}, the creation of strongly correlated phases through three body losses~\cite{Syassen2008strong}, and $N$-body Efimov states~\cite{Sorensen2002}.

Experimental investigation of ultracold bosons near unitarity has been hampered by the fast increase of the three-body recombination rate close to a Feshbach resonance \cite{KetterleFeshbach,roberts2000magnetic}. In this case, the number of trapped atoms $N(t)$ follows the usual three-body law
\begin{equation}
\dot N={-}L_3\langle n^2\rangle N,
\label{Eq1}
\end{equation}
\noindent where $\langle n^2\rangle\,{=}\,\int d^3 r \, n^3(\bm r)/N$ is the mean square density and $L_3$ is the three-body loss rate constant. In the zero-temperature limit $L_{3}$ increases as $\hbar a^4/m$~\cite{weber2003three} multiplied by a dimensionless log-periodic function of $a$ revealing Efimov physics~\cite{nielsen1999low,EsryGreeneBurke,bedaque2000three,BraatenHammerPhysRep2006,PetrovLesHouches,kraemer2006evidence,zaccanti2009observation,pollack2009universality,KhaykovichEfimov,ferlaino2010forty}. At finite temperature, $L_3$ saturates when $a$ becomes comparable to the thermal wavelength $\lambda_{\rm th}\,{=}\,h/\sqrt{2\pi mk_{\rm B}T}$, and $L_3\,{\sim}\,\hbar a^4/m\,{\sim}\,\hbar^5/m^3 (k_B T)^2$ \cite{dincao2004limits,dincao2009shortrange,HoUnitaryBose}. This saturation suggests that a non-quantum-degenerate Bose gas near a Feshbach resonance will maintain thermal quasiequilibrium \cite{HoUnitaryBose}. Indeed, in this regime,  $|a|\,{\gtrsim}\,\lambda_{\rm th}$ and $n \lambda_{\rm th}^3{\ll}1$. Thus, the elastic collision rate $\gamma_2\,{\propto}\,\hbar\lambda_{\rm th}n/m$ is much higher than the three-body loss rate $\gamma_3\,{=}\,L_3 n^2\,{\propto}\,\hbar\lambda_{\rm th}^4 n^2 /m$. Experimental and numerical evidence for a saturation of $L_3$ were reported in \cite{kraemer2006evidence,wild2012measurements,dincao2004limits}. A theoretical upper bound compatible with this scaling was derived in \cite{mehta2009general} assuming that only the lowest 3-body hyperspherical harmonic contributes, an assumption which breaks down when $|a|$ exceeds $\lambda_{\rm th}$.

%In contrast to most previous experiments, we operate in close vicinity of a Feshbach resonance in the non-degenerate regime. Namely, we have $|a|~\gtrsim~\lambda_{th}$ and $\lambda_{th}n^3\ll 1$. We explore the idea \cite{HoUnitaryBose} that under these conditions the elastic collision rate $\gamma_2\propto \hbar\lambda_{th}n/gammam$ is much higher than the three-body loss rate $\gamma_3=L_3 n^2\propto \hbar\lambda_{th}^4n^2/m$, and the trapped gas is maintained in quasi-thermal equilibrium.

In this Letter, we measure the temperature dependence of the unitary three-body recombination rate and find agreement with a $L_3\,{\propto}\,1/T^2$ scaling law. In a second set of measurements performed at constant temperature we study $L_3$ versus $a$. We show how this function smoothly connects to the zero-temperature calculations when $|a|\,{\ll}\,\lambda_{\rm th}$. These observations are explained by a general theoretical result for $L_3(a,T)$, exact in the zero-range approximation, that we derive in the second part. Our theory allows for a complete analytic description of the unitary case and, in particular, predicts (weak) log-periodic oscillations of the quantity $L_3T^2$. Our findings quantify the ratio of good-to-bad collisions in the system and provide a solid ground for future studies of strongly interacting Bose gases. Furthermore, on the $a\,{<}\,0$ side, experiments have so far detected a single Efimov trimer~\cite{wild2012measurements,zaccanti2009observation,pollack2009universality,KhaykovichEfimov,Gross2010nuclear}. Our analysis predicts that a second Efimov trimer of very large size should be detectable in $^7$Li at temperatures on the order of a few microkelvins.
%Interestingly, at
%constant $T$ we show theoretically that $L_3(a)$ reaches its
%maximum not at unitarity but at a finite value of $a\,{<}\,0$, a
%phenomenon which we attribute to a thermally broadened three-body
%resonance associated with an Efimov trimer of very large size

Our experimental setup was presented in \cite{navon2011dynamics}. After magneto-optical trapping and evaporation in a Ioffe magnetic trap down to ${\simeq}\,30~\mu$K, ${\simeq}\,2{\times}10^6$ $^7$Li atoms are transferred into a hybrid magnetic and dipole trap in the state $|1,1\rangle$. The transverse confinement is obtained by a single laser beam of waist $43(1) \mu$m and wavelength 1073~nm, while the longitudinal trapping is enhanced by a magnetic field curvature. The resulting  potential has a cylindrical symmetry around the propagation axis of the laser and is characterized by trapping frequencies $0.87\,\rm{kHz}\,{<}\,\omega_\rho/2\pi\,{<}\,3.07\rm\,{kHz}$ and $18\,\rm{Hz}\,{<}\,\omega_z/2\pi\,{<}\,49\,\rm{Hz}$. Further cooling is achieved by applying a homogeneous magnetic field $B\,{\simeq}\,718$~G for which the scattering length is ${\simeq}\,200a_0$, and decreasing the depth of the trapping potential down to a variable value $U'$ allowing us to vary the final temperature of the cloud. Afterwards, the dipole trap is recompressed to a value $U\,{>}\,U'$, to prevent significant atom loss due to the enhanced evaporation rate, see below. At each $T$ we choose $U$ so as to maintain the temperature constant during the three-body loss rate measurement. Finally, the magnetic field is ramped in $100{-}500$~ms to $B_{0}\,{\simeq}\,737.8(3)$~G where the scattering length $a$ diverges \cite{navon2011dynamics}.  We then measure the total atom number $N$ remaining after a variable waiting time $t$ and the corresponding $T$, using \emph{in situ} imaging of the thermal gas.

\begin{figure}
\centerline{\includegraphics[width=\columnwidth]{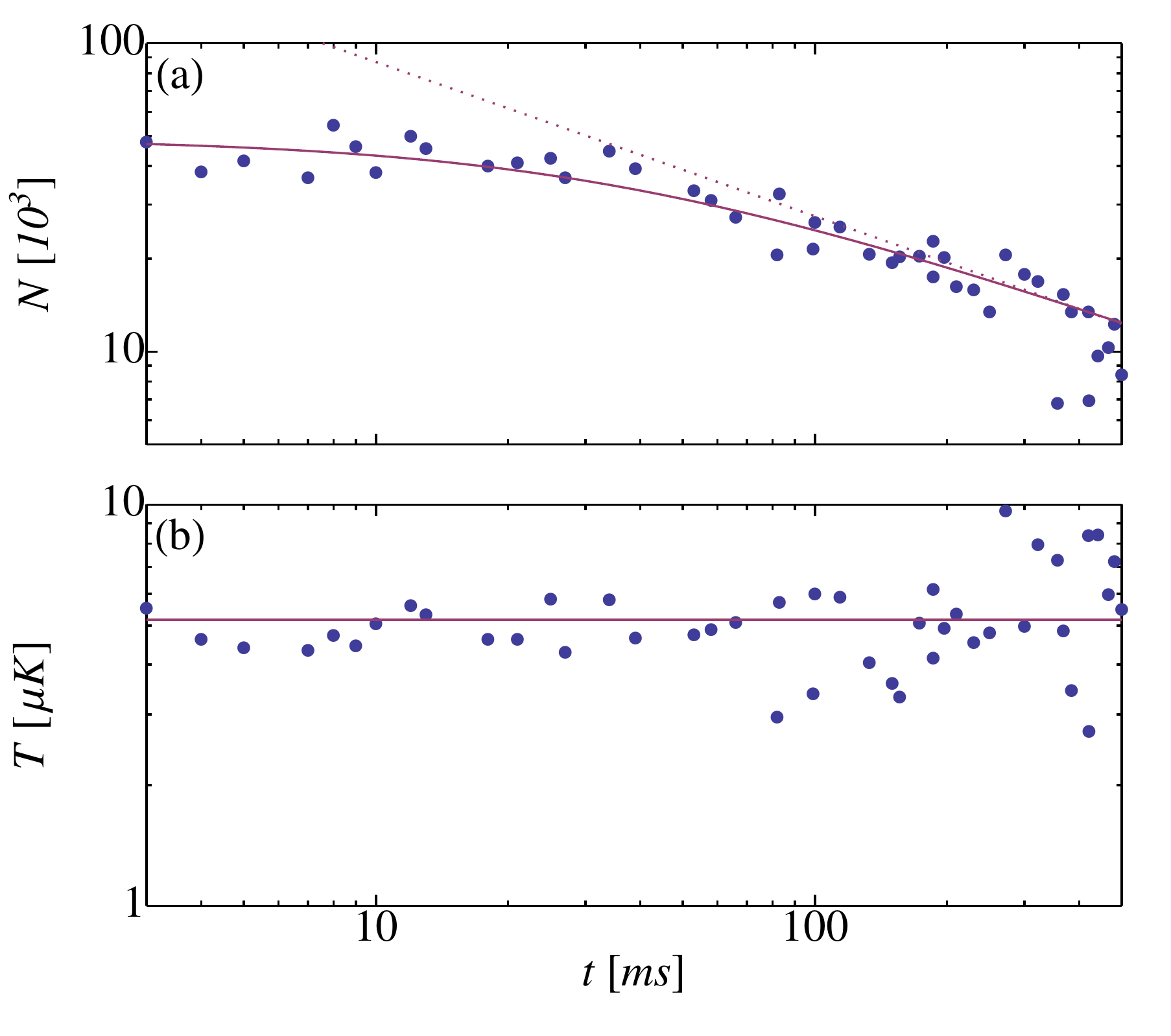}}
\caption{Time dependence of the atom number (a) and temperature (b) for $U=\eta \, k_{\rm B}T$, with $T = 5.2(4)~\mu {\rm K}$, $\eta  = 7.4$ and (uncorrected) $L_{3} = 1.2(2)_{\rm stat}\times10^{-21} {\rm cm}^{6} {\rm s}^{-1}$. The dotted line shows the long time $t^{-1/2}$ dependence of the number of atoms.}
%Ref. 8:20111110c
\label{fig:NTvst}
\end{figure}

Our data are limited to the range of temperature $1 \mu{\rm K}\,{\leq}\,T\,{\leq}\,10\mu{\rm K}$. For $T\,{\gtrsim}\,1\mu$K, the rate $\gamma_3\,{=}\,{-}\dot N/N$ remains small with respect to other characteristic rates in our cloud (elastic scattering rate, trapping frequencies), which guarantees that a thermal quasiequilibrium is maintained. We check that for these parameters the {\it in situ} integrated density profile is indeed gaussian, and we use it to extract  the temperature of the cloud, found to be in agreement with that of time of flight. The peak phase-space density varies within $0.07\,{\times}\,10^{-2}\,{<}\,n_{0} \lambda_{\rm th}^{3} \,{<}\,1.1\,{\times}\,10^{-2}$. A typical time dependence of $N$ and $T$ is shown in Fig.~\ref{fig:NTvst}. The time dependence of the atom number is fitted using the usual three-body recombination law Eq.~\eqref{Eq1} \cite{note1}. For a non-degenerate gas of temperature $T$, the density profile is gaussian, and we have $\langle n^2\rangle\,{=}\,N^{2} A(T)\,{=}\,N^2 (m\bar\omega^2/ 2\pi \sqrt{3} k_{\rm B} T)^3$, with $\bar\omega\,{=}\,(\omega_\rho^{2} \omega_z)^{1/3}$ being the mean trapping frequency. We then have:
\begin{equation}
\dot N=-L_{3} (T) A(T) N^3.
\label{Eq2}
\end{equation} 
Assuming constant temperature, integrating Eq.~\eqref{Eq2} gives
\begin{equation}
N(t)=\frac{N(0)}{\sqrt{1+2 A(T)L_{3}(T)N^2(0)t}},
\label{Eq3}
\end{equation}
which we use as a fitting function to analyze $N(t)$, and extract $L_3 (T)$ as shown in Fig.~\ref{fig:NTvst}.

Because of their $n^{3}/T^2$ dependence, three-body losses preferentially remove atoms of low kinetic energy and those located at the center of the trap where the density is the highest and potential energy is the smallest. As a result, three-body loss events heat up the cloud \cite{weber2003three}. We ensure constant temperature by operating with a typical trap depth $U\,{\simeq}\,\eta k_{\rm B}T$ with $6\,{\leq}\,\eta\,{\leq}\,8$, for which the residual evaporation then balances recombination heating, see Fig.~\ref{fig:NTvst}b. This ensures that $L_3$ is time independent, but, as a drawback, evaporation contributes to losses. To quantify the relative importance of evaporative and three-body losses, we first note that an atom expelled by evaporation removes on average an energy ${\simeq}\,( \eta + \kappa)\,k_{\rm B}T$, where, taking $\kappa$ from~\cite{luiten1996}, we follow~\cite{luo2006evap}. Typically, we have $\kappa\,{\simeq}\,0.68$ for $\eta\,{=}\,6$ and $\kappa\,{\simeq}\,0.78$ for $\eta\,{=}\,8$~\cite{SupplMat}. In comparison, each three-body event leaves on average an excess heat of $\delta k_{\rm B} T$ per particle. Extending the derivation of  \cite{weber2003three} to the case of an energy dependent three-body loss rate ${\propto}\,E^{-2}$, we obtain $\delta\,{=}\,5/3$ \cite{SupplMat}. The energy balance required to keep the temperature constant thus implies that the evaporation rate is ${\simeq}\,\delta / (\eta+\kappa-3)$ times smaller than the three-body loss rate. Neglecting this effect would induce a systematic overestimation of $L_3$ of about $50\,\%$ for $\eta\,{=}\,6$ and $30\,\%$ for $\eta\,{=}\,8$. Therefore, we apply this systematic correction to our data.

\begin{figure}
\centerline{\includegraphics[width=\columnwidth]{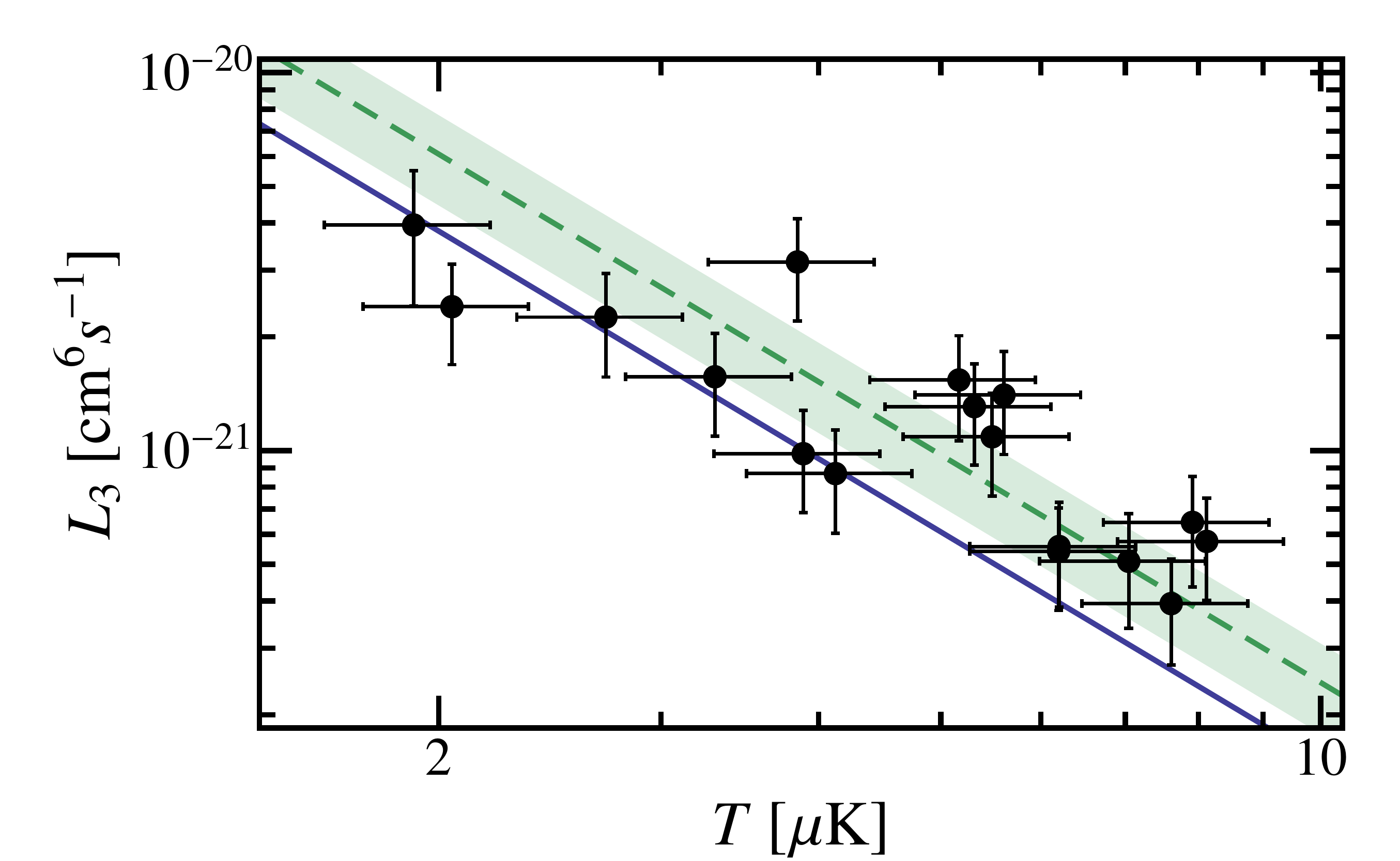}}
\caption{Temperature dependence of the three-body loss rate $L_3$.
Filled circles: experimental data; green dashed line: best fit to the data $L_3(T)\,{=}\,\lambda_{3}/T^2$
with $\lambda_{3}\,{=}\,\lambdafull$; the green band shows the  $1 \sigma $ quadrature sum of uncertainties. Solid line: prediction from Eq.~\eqref{eq:L3approx}, $\lambda_{3}\,{=}\,\lambdatheo$ with $\eta_*\,{=}\,0.21$ from \cite{Gross2011,Gross2010nuclear}.} \label{fig:L3vsT}
\end{figure}

The temperature dependence of $L_3$ obtained from our measurements at unitarity is shown in Fig.~\ref{fig:L3vsT}. It is well fit by the scaling law $L_3 (T)\,{=}\,\lambda_{3}/T^2$, with $\lambda_{3}\,{=}\,\lambdastat$ as the best-fit value. In order to discuss the systematic uncertainty of this measurement we note that the quantity $L_{3}T^{2}$ scales in all experimental parameters identically to the thermodynamic quantity $(\mu^{2}/P)^{2}$ of a zero-temperature BEC, with chemical potential $\mu$, and pressure $P$ \cite{SupplMat}. We use this relation to calibrate our experimental parameters \cite{navon2011dynamics} and obtain a systematic uncertainty on $\lambda_{3}$ of ${\leq}\,25\,\%$ resulting in $\lambda_{3}\,{=}\,\lambdafull$.

\begin{figure}
\centerline{\includegraphics[width=\columnwidth]{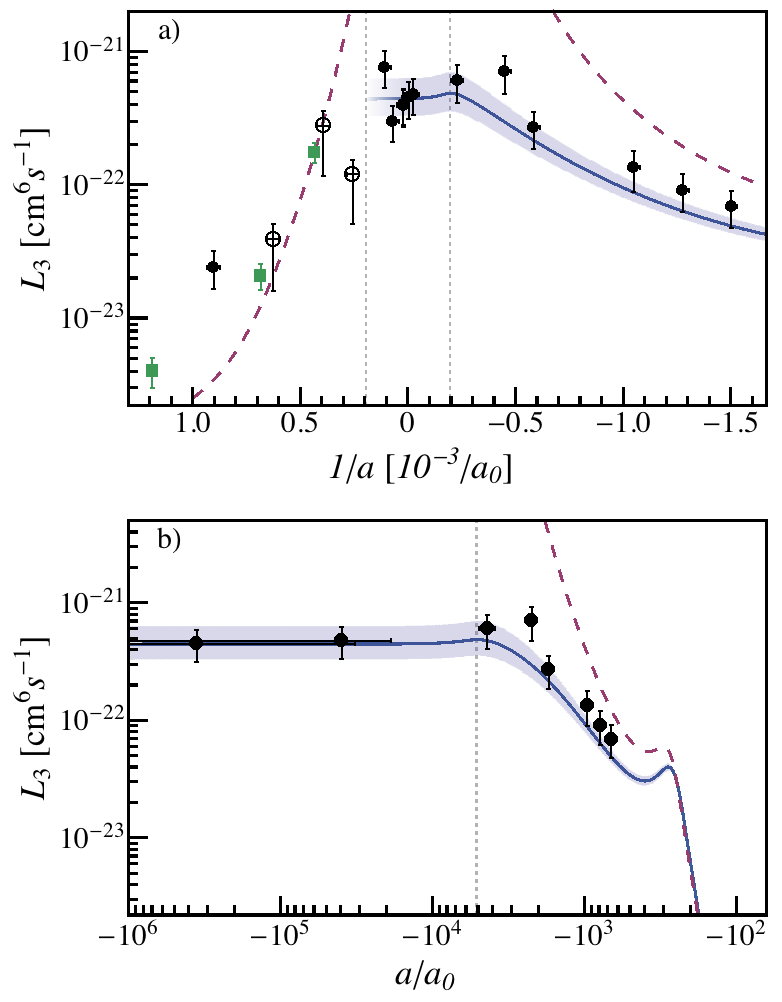}}
\caption{a) $^7$Li scattering-length dependence of the three-body rate constant $L_3(a)$ for constant $T\,{=}\,5.9(6)~\mu$K (filled and open circles). For small positive $a$, $L_3(a)$ for a low temperature condensate is also shown (green squares). The solid blue line corresponds to our theoretical prediction Eq.~\eqref{L3result} for $T\,{=}\,5.9~\mu$K. The blue range is the same theory for 5.3 to 6.5~$\mu$K. The dashed lines show the zero-temperature prediction for $L_3(a)$~\cite{BraatenHammerPhysRep2006} fitted to the measurements in~\cite{Gross2011,Gross2010nuclear} with the parameters $\eta_{*}\,{=}\,0.21$ and $R_0\,{=}\,270\,a_0$. The vertical dotted lines correspond to $|a|/\lambda_{\rm th}\,{=}\,1$. The open circles in the range $1500\,a_{0}\,{<}\,a\,{<}\,5000\,a_{0}$ are not corrected for residual evaporation as our model is not applicable. b) Logarithmic plot of the $a\,{<\,}0$ side, displaying the two Efimov loss resonances.}
\label{fig:L3vsa}
\end{figure}

We now study the $a$-dependence of $L_3$ on both sides of the resonance by employing the same experimental procedure as in the unitary case. We tune the scattering length while keeping the temperature within 10\,\% of $5.9~\mu$K, see Fig.~\ref{fig:L3vsa}. The excess heat $\delta$ entering in the correction now depends on the value of $ka$. The correction is applied to all data points (filled circles) except in the range $1500\,a_{0}\,{<}\,a\,{<}\,5000\,a_{0}$ (open circles), where the assumptions of our model are not applicable \cite{SupplMat}. In the limit $|a|\,{\gg}\,\lambda_{\rm th}$, we observe that $L_3(a)$ saturates to the same value on both sides of the resonance. In the opposite limit $|a|\,{\ll}\,\lambda_{\rm th}$, our data connect to the zero temperature behavior \cite{BraatenHammerPhysRep2006} studied experimentally in \cite{kraemer2006evidence,zaccanti2009observation,pollack2009universality,KhaykovichEfimov,ferlaino2010forty}. On the $a\,{<}\,0$ side, the dashed line is the zero-temperature prediction for $L_{3}$ from \cite{BraatenHammerPhysRep2006}. We clearly see that finite temperature reduces the three-body loss rate. On the $a\,{>}\,0$ side, temperature effects become negligible for $a\,{<}\,2000\,a_0$  as testified by our measurements performed on a low temperature Bose-Einstein condensate (green squares) which agree with the total recombination rate to shallow and deep dimers calculated at $T\,{=}\,0$ in \cite{BraatenHammerPhysRep2006} (dashed line). The data around unitarity and on the $a\,{<}\,0$ side are seen to be in excellent agreement with our theory Eq.~\eqref{L3result} described below.

In order to understand the dependence $L_3(a,T)$ theoretically, we employ the $S$-matrix formalism developed in~\cite{Efimov1979,BraatenHammerPhysRep2006,braaten2008three}. According to the method, at hyperradii $R\,{\gg}\,|a|$ one defines three-atom scattering channels ($i\,{=}\,3,4,...$) for which the wavefunction factorizes into a normalized hyperangular part, $\Phi_i(\hat{R})$, and a linear superposition of the incoming, $R^{-5/2}e^{-ikR}$, and outgoing, $R^{-5/2}e^{+ikR}$, hyperradial waves. The channel $i\,{=}\,2$ is defined for $a\,{>}\,0$ and describes the motion of an atom relative to a shallow dimer. The recombination or relaxation to deep molecular states (with a size of order the van der Waals range $R_e$) requires inclusion of other atom-dimer channels. In the zero-range approximation, valid when $R_e\,{\ll}\,R_{m}\,{\equiv}\,\rm{Min}\left( 1/k,|a|\right)$, the overall effect of these channels and all short-range physics in general can be taken into account by introducing a single Efimov channel ($i\,{=}\,1$) defined for $R_e\,{\ll}\,R\,{\ll}\,R_{m}$: the wavefunction at these distances is a linear superposition of the incoming, $\Phi_1(\hat{R})R^{-2+is_0}$, and outgoing, $\Phi_1(\hat{R})R^{-2-is_0}$, Efimov radial waves. Here $s_0\,{\approx}\,1.00624$. The notion ``incoming'' or ``outgoing'' is defined with respect to the {\it long-distance} region $R_{m}\,{\lesssim}\,R\,{\lesssim}\,|a|$, so that, for example, the incoming Efimov wave actually propagates towards larger $R$ whereas incoming waves in all other channels propagate towards smaller hyperradii. The matrix $s_{ij}$ relates the incoming amplitude in the $i^{\rm th}$ channel with the outgoing one in the $j^{\rm th}$ channel and describes the reflection, transmission, and mixing of channels in the {\it long-distance} region. This matrix is unitary and independent of the short-range physics. The short-range effects are taken into account by fixing the relative phase and amplitude of the incoming and outgoing Efimov waves $R^2\Psi\,{\propto}\,(R/R_0)^{is_0}{-}e^{2\eta_*}(R/R_0)^{-is_0}$, where $R_0$ is the three-body parameter and the short-range inelastic processes are parametrized by $\eta_*\,{>}\,0$, which implies that the number of triples going towards the region of $R\,{\sim}\,R_e$ is by the factor $e^{4\eta_*}$ larger than the number of triples leaving this region \cite{Braaten2003}. Braaten {\it et al.} \cite{braaten2008three} have shown that for a given incoming channel $i\,{\geq}\,2$ the probability of recombination to deeply bound states is $P_i\,{=}\,(1-e^{-4\eta_*})|s_{i1}|^2/|1+(kR_0)^{-2is_0}e^{-2\eta_*}s_{11}|^2$ \cite{Remarks11}. 
%We assume that our gas is in thermal equilibrium and all these channels are equally populated. 
For $a\,{<}\,0$, by using the fact that $s_{11}$ is unitary ($\sum_{i=1}^\infty|s_{1i}|^2\,{=}\,1$) and averaging over the Boltzmann distribution we then obtain the total loss rate constant 
%\begin{equation}\label{L3result}
%L_3=\frac{144\sqrt{3}\pi^2\hbar\sinh (2\eta_*)}{mk_{\rm th}^{6}e^{2\eta_*}}\int_0^\infty\frac{(1-|s_{11}|^2)e^{-k^2/k_{\rm th}^2}kdk}{|1+(kR_0)^{-2is_0}e^{-2\eta_*}s_{11}|^2},
%\end{equation}
\begin{equation}\label{L3result}
L_3\,{=}\,\frac{72\sqrt{3}\pi^2\hbar (1{-}e^{-4\eta_{*}})}{mk_{\rm th}^{6}}\!\int_0^\infty\!\!\frac{(1{-}|s_{11}|^2)e^{-k^2/k_{\rm th}^2}kdk}{|1{+}(kR_0)^{-2is_0}e^{-2\eta_*}s_{11}|^2},
\end{equation}
where $k_{\rm th}\,{=}\,\sqrt{mk_BT}/\hbar$.

Note that in deriving Eq.~\eqref{L3result} we closely followed \cite{braaten2008three} where the scattering length was assumed to be finite. However, we easily generalize this derivation to the case $a\,{=}\,\infty$, in which the channels become decoupled at distances $R\,{\gg}\,1/k$ and the {\it long-distance} region can now be defined by $R\,{\sim}\,1/k$. A less trivial result of our analysis is that for any $ka$ there exists a unitary transformation of the matrix $s_{ij}$ which leaves the element $s_{11}$ invariant, but all channels with $i\,{>}\,3$ become decoupled from the Efimov channel \cite{SupplMat}. This transformation constructs a new large-$R$ channel characterized by a certain hyperangular wavefunction $\tilde{\Phi}_3(\hat{R})$. For negative or infinite $a$ this is the only channel that can ``talk'' to the lossy short-distance Efimov channel via a unitary $2\times2$ matrix. Therefore, the three-body loss rate can not exceed the so-called maximum value $L_{3}^{\rm max}\,{=}\,36\sqrt{3}\pi^2\hbar^5(k_{\rm B} T)^{-2}/m^3$ reached in the case when the outgoing flux in this newly constructed channel vanishes. Previous derivations of $L_3^{\rm max}$ \cite{mehta2009general} essentially implied that $\tilde\Phi_3(\hat{R})$ is the lowest {\it non-interacting} hyperspherical harmonics. This approximation can be made only for $k|a|\,{\ll}\,1$. In general, $\tilde{\Phi}_3(\hat{R})$ is {\it not} an eigenstate of the angular momentum operator. In particular, at unitarity $\tilde{\Phi}_3(\hat{R})\,{=}\,\Phi_1(\hat{R})$ \cite{SupplMat}.

The function $s_{11}(ka)$ is calculated in \cite{SupplMat}. At unitarity it equals $s_{11}(\infty)\,{=}\,{-}e^{-\pi s_0}e^{2 i [s_0 \ln 2+ {\rm Arg}\, \Gamma(1+i s_0)]}$ and from Eq.~\eqref{L3result} one sees that $L_3T^2$ should be a log-periodic function of $T$. However, due to the numerically small value of $|s_{11}|\,{\approx}\,0.04$, in the case of three identical bosons the oscillations are very small and $L_3$ is well approximated by setting $s_{11}\,{=}\,0$:
\begin{equation}\label{eq:L3approx}
L_3\approx \frac{\hbar^5}{m^{3}}\,36\sqrt{3}\,\pi^2 \frac{1-e^{-4\eta_*}}{(k_B T)^{2}}.
\end{equation}
This explains the $L_3{\propto}T^{-2}$ experimental observation seen in Fig.~\ref{fig:L3vsT} at unitarity. Taking  $\eta_{*}\,{=}\,0.21$, which is the average of two measurements made for our $^7{\rm Li}$ Feshbach resonance in~\cite{Gross2011,Gross2010nuclear}, we get $L_3\,{=}\,\lambda_3/T^{2}$ with $\lambda_3\,{=}\,\lambdatheo$. This is $40\,\%$ below the experimentally determined value without any adjustable parameter and the agreement between theory and experiment is $1.4 \sigma$.
%Our data is also consisted with the upper bound in \ref{}.
%and is satisfactory considering that neglecting the (small)
%residual evaporation from the trap in our analysis overestimates
%$L_3$ by typically $30\%$.
We should point out that Eq.~\eqref{L3result} can be easily generalized to the case of other three-body systems with smaller $s_0$. Then, the terms neglected in Eq.~\eqref{eq:L3approx} can become important. They also become important in our system of three identical bosons when departing from resonance in the direction of $a\,{<}\,0$. Then $|s_{11}(ka)|$ monotonically increases as a function of $1/k|a|$ reaching 1 in the limit $ka{\rightarrow}0^{-}$, the argument of $s_{11}$ also being a monotonic function of $1/k|a|$ \cite{SupplMat}. The solid blue line in Fig.~\ref{fig:L3vsa} is the result obtained from Eq.~\eqref{L3result} using the same $\eta_*$ as above and $R_0\,{=}\,270 a_0$ also taken from \cite{Gross2011,Gross2010nuclear}. The shaded blue area reflects our experimental range of temperatures. More or less visible maxima of $L_3$ appear when the denominator in the integrand of Eq.~\eqref{L3result} reaches its minimum, {\it i.e.}, becomes resonant. The approximate condition for this is ${\rm Arg}\,s_{11}(ka)\,{=}\,\pi{+}2s_0\ln kR_0$ and the features become increasingly more pronounced for larger $|s_{11}|$ and smaller $\eta_*$. Note that from the viewpoint of the visibility of the maxima, decreasing $|a|$ is equivalent to decreasing $\sqrt{T}$. Fig~\ref{fig:L3vsa}b shows the pronounced resonance at $a\,{=}\,a_{-}{\approx}-274 a_0$ observed in~\cite{Gross2011,Gross2010nuclear}. This resonance is associated with the passage of an Efimov trimer through the three-atom threshold. Another Efimov trimer, larger in size by a factor of $e^{\pi/s_0}\,{=}\,22.7$, is expected to go through the threshold at around $a\,{\approx}\,{-}6350 a_0$ leading to another zero energy resonance. As we deduce from Eq.~\eqref{L3result} and show in Fig.~\ref{fig:L3vsa} for $5.9 \mu{\rm K}$, the thermally averaged remnants of this predicted resonance lead to a maximum of $L_3$ at $a\,{\approx}\,{-}5100 a_0$. As seen in Fig.~\ref{fig:L3vsa}b the agreement between theory and experiment is very good over the entire $a\,{<}\,0$ range.

%With our error bars we can not claim that we see this feature experimentally. However, on a larger scale, the theoretical prediction \benno{agrees with} the experimental data over the whole $a{<}0$ range.
% although with a slight vertical offset,
%most probably due to the uncertainty related to the residual
%evaporation.

Because of the existence of a shallow dimer state, the case $a\,{>}\,0$ becomes, in general, a complicated dynamical problem which should take into account the atom-dimer and dimer-dimer relaxation as well as various non-universal factors: the finite trap depth, chemical imbalance between trapped shallow dimers and free atoms, and deviations from thermal equilibrium which possibly depend on the preparation sequence. These issues require an extensive discussion beyond the scope of this Letter. The situation obviously simplifies in the case of very small $a$ when the system is purely atomic and the three-body recombination to deep and shallow molecules leads to an immediate loss of three atoms. 

Discussing the opposite limit of large $a\,{>}\,0$, we first note that dimers are well defined when their size $\,{\sim}\,a$ is smaller than $n^{-1/3}$, which we assume in the following (the limit $n a^3\,{\gg}\,1$ is equivalent to the case $a\,{=}\,\infty$). In the regime $a\,{\gg}\,\lambda_{\rm th}$ we find using the Skorniakov-Ter-Martirosian equation that $s_{12}{\to}0$ for $ka{\to}\infty$, which implies that the atom-dimer relaxation rate vanishes; shallow dimers then remain at chemical quasi-equilibrium with the decaying atomic ensemble, with a molecular fraction $\propto n \lambda_{\rm th}^3 \ll 1$ (for the data of Fig. \ref{fig:L3vsa} with $a\,{>}\,\lambda_{\rm th}$, we find a fraction of $0.6\%$)~\cite{SupplMat}.
%which implies that the atom-dimer relaxation rate vanishes and that shallow dimers remain at chemical quasi-equilibrium with the decaying atomic ensemble~\cite{SupplMat}. 
Shallow dimer formation and breakup are then balanced, so that the atomic decay is just given by Eq.~\eqref{Eq1}. The expression of $L_3$ for $a\,{>}\,0$ was obtained in~\cite{braaten2008three} and reduces to Eq.~\eqref{L3result} for $s_{12}\,{\to}\,0$. We conclude that the loss rate must be continuous across the resonance, in accordance with our experimental data. Therefore, in Fig.~\ref{fig:L3vsa}a the result of Eq.~\eqref{L3result} is simply continued to positive $a$ for $a\,{\gg}\,\lambda_{\rm th}$.

In summary, we have systematically studied the dependence of the three-body loss rate on $T$ and $a$ in a Bose gas near unitarity. Eq. \eqref{eq:L3approx} shows that, at unitarity, $L_{3}$ never reaches $L_{3}^{\rm max}$ and one can hope to produce quantum degeneracy in a unitary Bose gas using atomic species with a particularly small $\eta_{*}$. Note that the loss mechanism in our system drastically differs from a chemical reaction with finite activation energy $\Delta E$ characterized by the well-known Arrhenius law  $L_3\,{\propto}\,\exp(-\Delta E/k_B T)$. In our case, instead of a potential hill there is an effective three-body $R^{-2}$-attraction leading to $\Psi(R)\,{\propto}\,(\lambda_{\rm th}/R)^2$ at distances $R_e\,{\lesssim}\,R\,{\lesssim}\,\lambda_{\rm th}$, where we normalized the three-body wavefunction $\Psi$ to unit volume and omitted its log-periodic $R$-dependence. We clearly see that the probability of finding three atoms in the recombination region is enhanced at small temperatures and scales as $|\Psi|^2\,{\propto}\, \lambda_{\rm th}^4\,{\propto}\,1/T^2$. More subtle is a quantum interference effect in Efimov three-body scattering, which leads to
%The $T^{-2}$ temperature dependence seems to contradict the well-known Arrhenius law  $L_3\,{\propto}\,\exp(-\Delta E/k_B T)$ for a chemical reaction with activation energy $\Delta E$. This paradox is resolved by noting that three-body recombination occurs through the ${-}1/R^{2}$ Efimov three-body potential that does not display any potential hill. The $T^{-2}$ behavior follows from the overlap of the incoming plane wave with the Efimov hyperangular wavefunction.
%Accordingly, the rate of the inelastic process is proportional to the probability of finding three bosons at distances $\sim \lambda_{\rm th}$ multiplied by the typical rate scale for these distances $\hbar/m\lambda_{\rm th}^2$ leading to the $T^{-2}$ law.
%~\cite{SupplMat}.
%The $T^{-2}$ behavior follows from the overlap of the incoming plane wave with the Efimov hyperangular  wavefunction~\cite{SupplMat}. 
%Moreover, a quantum interference effect in Efimov three-body scattering leads to 
an enhanced decay rate at a negative $a$, suggesting the possibility to observe the signature of a second Efimov trimer of large size. Another future direction is to explore the approach to the quantum degenerate regime and test whether the virial expansion of the unitary Bose gas~\cite{Castin2013} can be measured by using quasi-equilibrium thermodynamics \cite{HoUnitaryBose}.

\begin{acknowledgements}
We acknowledge fruitful discussions with G. Shlyapnikov and F. Ferlaino, and support from R\'egion \^Ile de France (IFRAF), EU (ERC advanced grant Ferlodim), Institut Universitaire de France, and the Russian Foundation for Fundamental Research. T.L. acknowledges support by the Austrian Science Fund (FWF) through the Doctoral Programme CoQuS (W1210).
\end{acknowledgements}

\end{document}

\title{Supplemental Material: Lifetime of the Bose Gas with Resonant Interactions}

\author{B.~S.~Rem}
\author{A.~T.~Grier}
\author{I.~Ferrier-Barbut}
\author{U.~Eismann}
\author{T.~Langen}
\author{N.~Navon}
\author{L. Khaykovich}
\author{F.~Werner}
\author{D.~S.~Petrov}
\author{F.~Chevy}
\author{C.~Salomon}

\date{\today}

\maketitle

%\documentclass[prl,aps,twocolumn,floatfix]{revtex4}
%\usepackage{amsmath,amssymb}%,graphicx}
%\usepackage[pdftex]{graphicx}
%\begin{document}
%\title{Supplemental material: Lifetime of the Unitary Bose Gas}
%\maketitle

%\twocolumn[
% % \begin{@twocolumnfalse}
%\center{Supplemental material: Lifetime of the Unitary Bose Gas}
%%\end{@twocolumnfalse}
% \vskip 0.5cm ]

\renewcommand{\theequation}{S\arabic{equation}}
\renewcommand{\thefigure}{S\arabic{figure}}

\setcounter{equation}{0}
\setcounter{figure}{0}

\section{Calculation of $s_{11}(ka)$}\label{sec1}

The function $s_{11\!}(k a)$ has been calculated for $0~{<}~ka~\lesssim~10$ in \cite{Braaten2008}. Here we calculate it for $ka$ of any sign and magnitude including infinity. The three-body Schr\"odinger equation reads
\begin{equation}\label{Schr}
\left[-\nabla^2_{\bf R}+V_a({\bf R})-k^2\right]\Psi({\bf R})=0,
\end{equation}
where $V_a$ is the sum of binary interaction terms parametrized by $a$ and the six-dimensional vector ${\bf R}$ is $\{(2{\bf r}_3-{\bf r}_1-{\bf r}_2)/\sqrt{3},{\bf r}_1-{\bf r}_2\}$, where ${\bf r_1,r_2}$ and ${\bf r_3}$ are atomic coordinates, and we set $\hbar=m=1$. By definition the first row of matrix $s_{ij}$ gives us the solution of Eq.~\eqref{Schr} with the following asymptotes. For small $R$ we have 
%\renewcommand{\arraystretch}{2.}
%\begin{equation}\label{Asymptotes}
%\Psi\approx\left\{\begin{array}{lr}\Phi_1(\hat{R})\left[(kR)^{is_0}+s_{11}(kR)^{-is_0}\right]/\sqrt{2s_0}R^2,& R\rightarrow 0,\\
%s_{12}\Psi_2({\bf R})+\sum_{i=3}^{\infty}s_{1i}\Phi_i(\hat{R})\frac{\exp(ikR)}{\sqrt{2kR}R^2},&R\rightarrow \infty,\end{array}\right.
%\end{equation}
%\renewcommand{\arraystretch}{1}
%\frac{\exp(is_0\ln kR)+s_{11}\exp(-is_0\ln kR)}{\sqrt{2s_0}R^2}
\begin{equation}\label{SmallRasymptote}
\Psi({\bf R})\approx \Phi_1(\hat{R})\left[(kR)^{is_0}+s_{11}(kR)^{-is_0}\right]/\sqrt{2s_0}R^2.
\end{equation}
Note that our definition of $s_{11}$ differs from the one of \cite{Braaten2008} by the factor $-(ka)^{2is_0}e^{-2i\delta_0}$, where $\delta_0\approx 1.588$~\cite{Macek}. In the asymptotic region of large $R$ we have
\begin{equation}\label{LargeRasymptote}
\Psi({\bf R})\approx s_{12}\Psi_2({\bf R})+\sum_{i=3}^{\infty}s_{1i}\Phi_i(\hat{R})e^{ikR}/\sqrt{2kR}R^2.
\end{equation}
In Eqs.~(\ref{SmallRasymptote}-\ref{LargeRasymptote}) all $\Phi_i(\hat{R})$ are symmetrized and normalized and $\Psi_2({\bf R})$ is the symmetrized wavefunction of the atom-dimer relative outgoing motion normalized to a unit flux. Physically $\Psi({\bf R})$ describes the stationary flow of atoms which are injected at the origin and can either return back with amplitude $s_{11}$ (second term in the right hand side of Eq.~\eqref{SmallRasymptote}) or travel to infinity by using channels with $i\geq 3$.

Before explaining the numerical method of calculating $s_{11}(ka)$ let us discuss some properties of this function which can be derived analytically. In order to do this it is convenient to use the complex scaling of the Hamiltonian \cite{Moiseyev1998} and multiply (rotate in the complex plane) all spatial coordinates by the complex number $ke^{-i\pi/2}$, i.e., we introduce $\tilde{\bf R}={\bf R}ke^{-i\pi/2}$. Then the problem reduces to calculating properties of the bound trimer state with energy $E=-1$, interaction between the atoms being characterized by the imaginary scattering length $\tilde{a}=ka/i$. Applying the complex scaling to the asymptotes (\ref{SmallRasymptote}-\ref{LargeRasymptote}) we see that now the solution is constrained to decay at large distances and to have the short-distance asymptote $\propto \tilde{R}^{is_0}+s_{11}(ka)e^{\pi s_0}\tilde{R}^{-is_0}$. That $\tilde{a}$ is imaginary simplifies the task: if $\Psi$ is a solution for a given real value of $a$, then $\Psi^*$ is the solution for $a=-a$. This leads to the relation $s_{11}(-ka)=e^{-2\pi s_0}/s_{11}^*(ka)$, and it is thus sufficient to deal, for example, only with $ka<0$. Since our problem is the inverse to finding the Efimov spectrum versus the three-body parameter and $a$, the point $ka=\infty$ and its vicinity can be treated analytically: the wavefunction of an Efimov trimer at unitarity is proportional to $\Phi_1(\hat{R})\left[J_{is_0}(i\tilde{R})-e^{-\pi s_0}J_{-is_0}(i\tilde{R})\right]$ which gives $s_{11}(\infty)=-2^{2is_0}e^{-\pi s_0}\Gamma(1+is_0)/\Gamma(1-is_0)$, the result presented in the main text. Moreover, by using the known analytic formula for the shift of the trimer energy at small $1/a$ with a fixed three-body parameter \cite{CastinWernerTrimere} one obtains $s_{11}(ka\gg 1)\approx s_{11}(\infty)(1-Cs_0/ka)$, where \cite{CastinWernerTrimere}
\bea \label{eq:c}
C & = & \pi \, {\rm sinh}\left(\frac{s_0 \pi}{2}\right) \, {\rm tanh}(s_0 \pi) /
\bigg[ {\rm cosh}\left(\frac{s_0 \pi}{2}\right) \nonumber \\ 
& + &  \frac{s_0\,\pi}{2} {\rm sinh}\left(\frac{s_0 \pi}{2} \right) - \frac{4 \pi}{3 \sqrt{3}} {\rm cosh}\left(\frac{s_0 \pi}{6}\right) \bigg] \nonumber \\
& = & 2.112 671 6 \ldots
\eea
Finally, the three-body wavefunction in the limit of vanishing total energy has been studied in \cite{Macek} from which we obtain $s_{11}\approx (k|a|)^{2is_0}e^{-2i\delta_0}$ in the limit $ka\rightarrow 0^{-}$.

In order to calculate $s_{12}$ for arbitrary $ka$ let us introduce the reduced wavefunction $f({\bf r})$ defined by
\begin{equation}\label{f}
f\left[(2{\bf r}_3-{\bf r}_1-{\bf r}_2)/\sqrt{3}\right]=4\pi \lim_{{\bf r}_1\rightarrow{\bf r}_2}|{\bf r}_1-{\bf r}_2|\Psi({\bf R}),
\end{equation} 
and write down the Skorniakov-Ter-Martirosian (STM) equation for the Fourier transform of $f({\bf r})$ (for more details see \cite{PetrovLesHouches})
\begin{equation}\label{STM}
(\sqrt{p^2-k^2}-1/a)f(p)-\hat{L}_{k^2}f(p)=0,
\end{equation}
where the integral operator $\hat{L}$ is defined by
\begin{equation}\label{hatL}
\hat{L}_{k^2}f(p)=\int_0^\infty \ln\left(\frac{p'^2+p^2+pp'-3k^2/4}{p'^2+p^2-pp'-3k^2/4}\right)\frac{4f(p')p'dp'}{\sqrt{3}\pi p}.
\end{equation}
%\begin{eqnarray}\label{STM}
%\!\!\!\!\!\!\!\!\!\!&&(\sqrt{p^2-k^2}-1/a)f(p)-\frac{4}{\sqrt{3}\pi p}\nonumber\\
%\!\!\!\!\!\!\!\!\!\!&&\times \int_0^\infty \ln\left(\frac{p'^2+p^2+pp'-3k^2/4}{p'^2+p^2-pp'-3k^2/4}\right)f(p')p'dp'=0.
%\end{eqnarray} 
Note that in Eq.~\eqref{STM} we use $f({\bf p})\equiv f(p)$ since, in the case of three identical bosons, higher spherical harmonics of this function correspond to the non-Efimovian kinematics, do not contribute to the asymptote~(\ref{SmallRasymptote}), and do not lead to (strong) recombination losses. 

As usual, the branches of the logarithm and of the square root are chosen as if the momentum $k$ (or energy $k^2$) is slightly shifted into the upper complex half-plane, or, alternatively, $p$ and $p'$ are slightly shifted to the lower half-plane. In fact, the complex scaling discussed above means that we rotate $p$ and $p'$ in the clockwise direction all the way to the negative imaginary axis and rescale them by $k$. Then changing variables in such a way that the integration goes along the positive real axis, we obtain the same Eq.~\eqref{STM} in which $k\rightarrow i$ and $1/a\rightarrow i/ka$. The resulting equation does not have singularities on the real axis and is extremely easy to solve numerically. The large-$p$ asymptote of the solution can be written as $C_1 p^{-2-is_0}+C_2 p^{-2+is_0}$, and it is straightforward to show that $s_{11}(ka)=(C_2/C_1)\left[\Gamma(1+is_0)/\Gamma(1-is_0)\right]e^{-\pi s_0}$.

\begin{figure}[hptb]
\begin{center}
\includegraphics[width=0.9\columnwidth,clip]{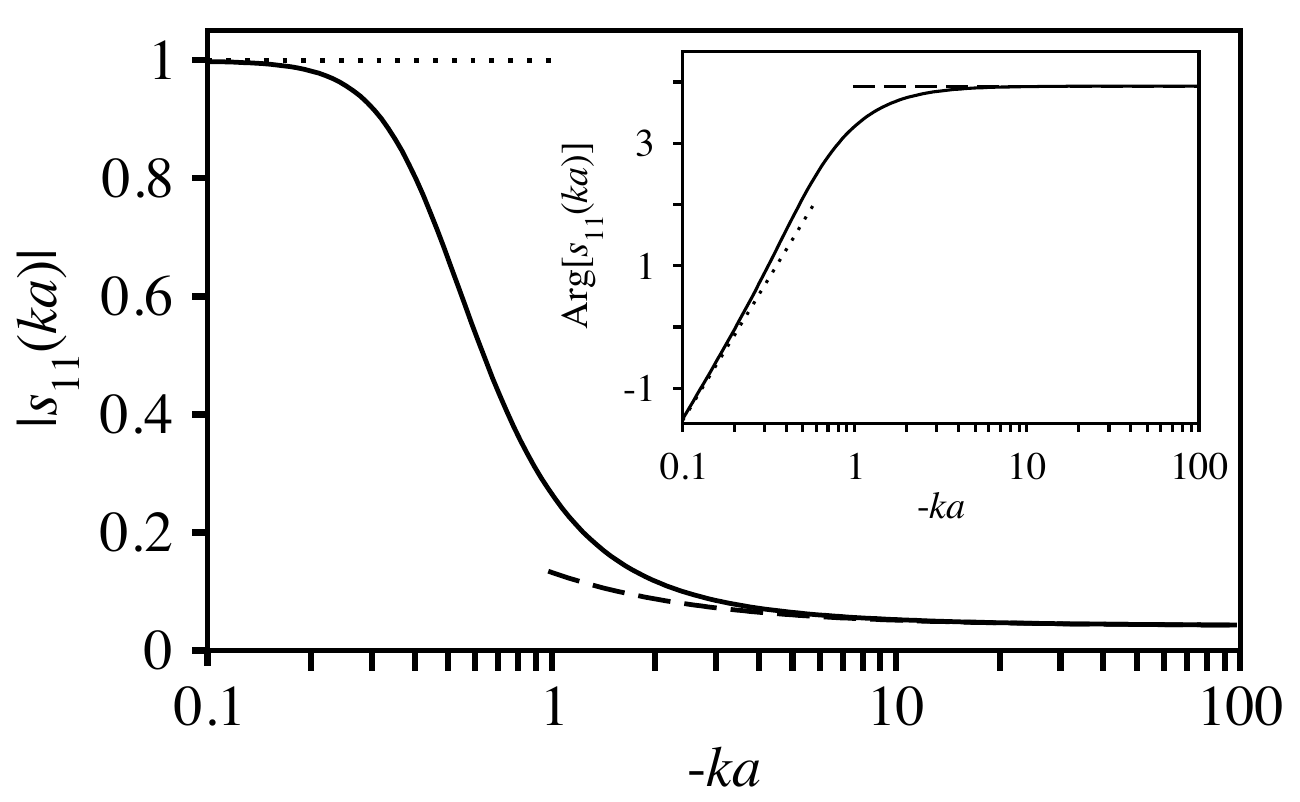}
\caption{Modulus and phase of $s_{11}$ versus $ka<0$. Dashed and dotted lines are analytic limits, see text.}
\label{fig:Figs11}
\end{center}
\end{figure}

In Fig.~\ref{fig:Figs11} we plot $|s_{11}|$ and ${\rm Arg}s_{11}$ versus $ka<0$ (solid line). The dashed and dotted lines correspond, respectively, to the limits $k|a|\gg 1$ and $k|a|\ll 1$ discussed above.
 
% We see that $|s_{11}|$
% remains
% small everywhere on the positive side of the resonance and close to the unitary point, %. We can then neglect it in Eq.~(\ref{L3result}) arriving at
% leading to
% the approximate plateau value Eq.~(\ref{eq:L3approx}) (see Fig.~\ref{fig:L3vsa}). Then departing from the resonance on the negative side $|s_{11}|$ increases, and tends to $1$
% in the limit $ka\rightarrow 0^-$, where
%  we have $s_{11}\approx (k|a|)^{2is_0}e^{-2i\delta_0}$
%   with $\delta_0\approx 1.588$ \cite{Macek} (see dotted line in Fig.~\ref{fig:Figs11})
%and we recover the zero-temperature result for the recombination rate constant (dashed line in Fig.~\ref{fig:L3vsa}). %Interestingly, for a given $k$ the integrand in Eq.~(\ref{L3result}) reaches its maximum when $s_{11\!}(ka)=-(kR_t)^{2is_0}e^{-2\eta}$, i.e. on the negative side of the resonance for our value of $\eta$. Although thermally broadened this maximum is still visible.

\section{Large $ka$ asymptote of  $|s_{12}(ka)|$}\label{sec2}

The quantity $s_{12}$ is the amplitude of the atom-dimer outgoing wave, see Eq.~\eqref{LargeRasymptote}. By construction it is zero for $a<0$ since there are no shallow dimers and, as we argue in the main text and in Sec. \ref{sec3}, for $a>0$ it becomes important for determination of $L_3$. The question that we address now is whether $|s_{12}|^2$ vanishes for $a \rightarrow +\infty$ or not. Unfortunately, numerical results for $s_{12}(ka)$ are available only for $ka\lesssim10$ \cite{Braaten2008} and do not allow us to make any statement on the large $ka$ behavior of this quantity.

In the region $|{\bf r}_1-{\bf r}_2|\sim a$ and $R\gg a$ the atom-dimer wavefunction $\Psi_2({\bf R})$ introduced in Eq.~\eqref{LargeRasymptote} can be written as
\begin{equation}\label{AtomDimerLargeHyperradius}
\Psi_2({\bf R})=\frac{\exp(-|{\bf r}_2-{\bf r}_1|/a)}{\sqrt{2\pi a}|{\bf r}_2-{\bf r}_1|}\frac{\exp(ip_0 R)}{\sqrt{24\pi p_0}R},
\end{equation}
where $p_0=\sqrt{k^2+1/a^2}>k$ is the atom-dimer relative momentum. The outgoing wave (\ref{AtomDimerLargeHyperradius}) corresponds to the pole of $f(p)$ at $p=p_0$:
\begin{equation}\label{AtomDimerPole}
f(p)\approx \frac{2\pi s_{12}(ka)}{\sqrt{3ap_0^3}}\frac{1}{p-p_0-i0},\; |p-p_0|\ll p_0-k. 
\end{equation}
Therefore, in order to calculate $s_{12}$ one has to solve Eq.~\eqref{STM} (with the correct boundary condition at large $p$) and find the residue of this pole. Here we solve this problem perturbatively using $1/ka$ as a small parameter.

The solution of the three-body problem at unitarity is given in terms of the Bessel functions. The corresponding correctly normalized function $f_0(p)$ up to a phase factor can be written as
\begin{eqnarray}\label{FourierBessel}
\!\!\!\!\!\!\!\!\!\!\!\!&&f_0(p)=\frac{2\pi\sqrt{Cs_0\cosh(\pi s_0)}\exp(-\pi s_0)}{\sqrt{3}\sinh(\pi s_0/2)p\sqrt{p^2-k^2}}\\
\!\!\!\!\!\!\!\!\!\!\!\!&&\times\left[k^{2is_0}e^{\pi s_0}(p+\sqrt{p^2-k^2})^{-is_0}-(p+\sqrt{p^2-k^2})^{is_0}\right]. \nonumber
\end{eqnarray} 
In deriving Eq.~\eqref{FourierBessel} we used the small $\hat{R}$ asymptote of the normalized hyperangular Efimov wavefunction $\Phi_1(\hat{R})$ which we took from \cite{CastinWernerTrimere}. The constant $C$ is defineed in Eq.~\eqref{eq:c}. 

Let us now write the solution of Eq.~\eqref{STM} at small $1/ka$ as $f(p)=f_0(p)+\delta f(p)$, where $\delta f(p)$ tends to zero when $a\rightarrow\infty$. Equation~(\ref{STM}) now reduces to
\begin{equation}\label{STMfordeltaf}
(\sqrt{p^2-k^2}-1/a)\delta f(p)=f_0(p)/a+\hat{L}_{k^2}\delta f(p).
\end{equation}
Looking at the right hand side of this equation at $p=p_0$ we observe that the first term tends to a finite value as $a\rightarrow \infty$ since $f_0(p)$ is singular at $p\rightarrow k$. In contrast, the integral operator smooths singularities and makes the second term vanish uniformly for large $a$. Therefore, the dominant contribution to $s_{12}$ can be obtained by neglecting the second term, and we finally obtain 
\begin{equation}\label{s12result}
|s_{12}(ka)|^2\approx 2Cs_0\left[1+\exp(-2\pi s_0)\right]/ka,\; ka\gg 1,
\end{equation} 
i.e., we have managed to show that $s_{12}\rightarrow 0$, as one approaches the resonance. 

\section{Atom-dimer chemical equilibrium near resonance}\label{sec3}

On the positive side of the resonance the loss rate in the system is no longer solely due to the recombination to deep molecular states. Three atoms can recombine to a shallow dimer and depending on how its binding energy, $E_D=\hbar^2/ma^2$, compares to the trap depth, $U$, the products of such a three-body event may or may not leave the trap. Moreover, even if they have enough energy to leave, they can collide with the remaining atoms and redistribute their excess energy into heat. This dynamical problem, in general, goes far beyond calculating the loss rate in three-atom collisions. Obviously, this complication is absent very far from the resonance where, starting from a purely atomic sample, one counts any recombination event as the loss of three atoms. By contrast, if $E_D<U$, the shallow dimers stay in the system and mix with atoms. 

We now focus on the regime $E_D\ll k_B T$, where the situation greatly simplifies. Let us assume chemical equilibrium between atoms and shallow dimers, and validate  this assumption {\it a posteriori}. The dimer density $n_D$ is then related to the atomic density $n$ by
%comparing the rates of elastic collisions with inelastic ones in the non-degenerate case, one can show that atomic and molecular components are in (quasi)thermal equilibrium with the same temperature. The question of the atom-dimer chemical equilibrium is more subtle and depends on how the rates of the following processes compare to each other: the recombination (relaxation) of atoms (shallow dimers) to deeply bound states, the recombination of atoms into shallow dimers and the inversed process -- break-up. For finite $\eta_*$ the rates of these processes can be comparable to each other. Therefore, we just admit that the molecular density $n_m$ can differ from its chemical equilibrium value $n_m^{(0)}$, which is related to the (purely) atomic density by
\begin{equation}\label{nD0}
n_D= n^2\lambda_{\rm th}^3\,2\sqrt{2}\,  e^{E_D/(k_B T)} \simeq n^2\lambda_{\rm th}^3\,2\sqrt{2} .
%(4\pi)^{3/2}n^2\kappa_T^{-3}e^{1/\kappa_T^2a^2}=n^2\left(\frac{4\pi\hbar^2}{mk_BT}\right)^{3/2}e^{E_D/k_BT}.
\end{equation}
The two reverse processes of three-atom recombination to a shallow dimer and of atom-dimer breakup (i.e. dissociation of a shallow dimer after collision with an atom) then balance each other and give a vanishing total contribution to $dn/dt$. Hence $dn/dt=-L_3\,n^3-L_2^{AD}\,n\,n_D$, where $L_{3}$ is the rate constant for recombination to deep dimers and $L_2^{AD}$ is the rate constant %three-atom recombination to a deeply bound dimer and 
for atom-dimer relaxation (i.e. formation of a deeply bound dimer after collision of a shallow dimer with an atom).
%The rate equation for the atomic density is given by Eq.(1) of the main text
%\begin{widetext}
%\begin{equation}\label{L3AAA}
%L_3(T)=\frac{36\sqrt{3}\pi^2\hbar^5[1-\exp(-4\eta_*)]}{(mk_B T)^3}\int_0^{\infty}\frac{1-|s_{11}|^2-|s_{12}|^2}{|1+(kR_0)^{-2is_0}e^{-2\eta_*}s_{11}|^2}e^{-E/k_BT}dE
%\end{equation}
The expression of $L_3(T)$ for $a>0$ was obtained in~\cite{Braaten2008} and differs from Eq.~(4) of the main text only by the replacement of the term $1-|s_{11}|^2$ by $1-|s_{11}|^2-|s_{12}|^2$. For the incoming atom-dimer channel $i\,{=}\,2$, the expression of the loss probability $P_2$ was given above Eq.~(4) in the main text, and leads after thermal averaging to
\begin{widetext}
\begin{equation}\label{L2AD}
L_2^{AD}(T)=\frac{3\sqrt{3\pi}\hbar^2[1-\exp(-4\eta_*)]}{2\,(mk_B T)^{3/2}}e^{-E_D/k_B T}\int_{-E_D}^{\infty}\frac{|s_{12}|^2}{|1+(|k|R_0)^{-2is_0}e^{-2\eta_*}s_{11}|^2}e^{-E/k_BT}dE.
\end{equation}
\end{widetext}
Here the integration variable $E=\hbar^2k^2/m$ is the total energy of the three-atom system in the center of mass reference frame. Thus, the integration over negative $E$ %in Eq.~(\ref{L2AD}) 
decribes the atom-dimer relaxation events below the break-up threshold. 
%In this case the free triatomic channels are closed, matrix $s_{ij}$ is a $2\times 2$ unitary matrix, and the quantity $1-|s_{11}|^2-|s_{12}|^2$ vanishes. In Eq.~(\ref{L3AAA}) we can then also set the lower integration limit to $-E_D$. 
We should also note that the matrix elements $s_{11}$ and $s_{12}$ are functions of $\sqrt{E}a$, which becomes imaginary for $E<0$. To show that $L_2^{AD}$ vanishes in the large $a$ limit, we treat separately the contributions from positive and negative $E$: For $E>0$, we have seen above that $s_{12}$ vanishes as $1/a$, see Eq.~\eqref{s12result}, which leads to a contribution $\propto 1/a$ to $L_2^{AD}$; for $E<0$, the integration is limited to the narrow window $[-E_D, 0]$, and the integrand can be bounded from above thanks to $P_2\leq1$, leading to a contribution $\propto1/a^2$ to $L_2^{AD}$. %, which is of higher order than the $E>0$ contribution.
This allows us to neglect $L_2^{AD}$ in the rate equation for $dn/dt$, which then reduces to Eq.~(1) of the main text~\cite{note4}.

Finally, let us validate our chemical equilibrium assumption. For a given dimer, the event rates for relaxation and breakup (after collision with an atom) are respectively $\gamma_{\rm rel} = n\,L_2^{AD}(T)$ and $\gamma_{\rm break}=n\,\alpha_{\rm shallow}(T)\,\lambda_{\rm th}^{-3}\,\sqrt{2}\,e^{-E_D/(k_B T)}$, where $\alpha_{\rm shallow}(T)$ is the event rate constant for three-atom recombination to a shallow dimer, and we used Eq. \eqref{nD0}. In the regime $a\gg\lambda_{\rm th}$ considered here, 
we can estimate from~\cite{Braaten2008} that
$3\alpha_{\rm shallow}(T)$ saturates to
a value $\gtrsim10\,L_3^{\rm max}(T)$~\cite{note5}.
Evaluating the leading-order behavior of $L_2^{AD}$ as explained above
then gives
 $\gamma_{\rm rel}/\gamma_{\rm break}  \lesssim 0.03 \, \lambda_{\rm th}/a \ll1$~\cite{note6}.
Hence
the relaxation  events do not destroy chemical equilibrium, as they happen much less frequently than the breakup events (and thus also than the reverse dimer-formation events). The relaxation rates in dimer-dimer and dimer-atom-atom collisions are also smaller than $\gamma_{\rm break}$ by factors $\propto n\lambda_{\rm th}^3 \ll1$ (with unknown prefactors which depend on the four-body problem). The last condition to check is $\gamma_{\rm break} \gg \gamma_3$, i.e., a given dimer should be likely to break up (and to be replaced by a newly formed dimer) within a time much smaller than the timescale $1/\gamma_3$ over which the cloud decays.
 Estimating $\gamma_{\rm break}$ as above and using our result for $\gamma_3$ gives $\gamma_3/\gamma_{\rm break} \lesssim 0.1\,n\lambda_{\rm th}^3\ll1$.

\section{Matrix $s_{ij}$: negative $a$}

In the case $a<0$ the atom-dimer channel is closed and the structure of matrix $s_{ij}$ is as follows. We have a single discrete small-$R$ Efimov channel and a continuum of large-$R$ channels, $\{\Phi_i(\hat{R}),\; i\geq 3\}$ being a complete orthonormal set of hyperangular functions, for example, eigenfunctions of the hyperangular kinetic energy operator in the absence of interactions. Given $s_{ij}$ we change this basis in favor of another orthonormal set $\{\tilde{\Phi}_i(\hat{R})\}$ in which we choose
\begin{equation}\label{Phi3Tilde}
\tilde{s}_{13} \tilde{\Phi}_3 = \sum_{i\geq 3} s_{1i} \Phi_i,
\end{equation}
The normalization condition which we impose on $\tilde{\Phi}_3$ uniquely defines this function and $\tilde{s}_{13}$ (up to an irrelevant phase factor). The corresponding asymptotic triatomic channel is defined by the incoming, $\psi_3=\tilde{\Phi}_3(\hat{R})e^{-ikR}/\sqrt{2kR}R^2$ and outgoing, $\psi_3^*$, waves. Note that we do not touch the Efimov channel and, therefore, $\tilde{s}_{11}=s_{11}$. From Eq.~\eqref{Phi3Tilde} and the unitarity of $s_{ij}$ we can deduce that $|\tilde{s}_{11}|^2+|\tilde{s}_{13}|^2=1$. Since the new matrix $\tilde{s}_{ij}$ should also be unitary, we conclude that $\tilde{s}_{11}$ and $\tilde{s}_{13}$ are the only non-zero entries of its first row. Let us now write explicitly the corresponding wavefunction [cf. Eqs.~(\ref{SmallRasymptote}-\ref{LargeRasymptote})]
\begin{equation}\label{Psi}
\Psi=\left\{\begin{array}{lr}\psi_1+\tilde{s}_{11}\psi_1^*,& R\rightarrow 0,\\
\tilde{s}_{13}\psi_3^*,&R\rightarrow \infty,\end{array}\right.
\end{equation}
where we denote the incoming Efimov wave as $\psi_1=\Phi_1(\hat{R})(kR)^{is_0}/\sqrt{2s_0}R^2$.

Choosing an appropriate linear combination of $\Psi$ and $\Psi^*$ we obtain the relation
\begin{equation}\label{PsiPlusPsi}
\frac{\Psi^*-\tilde{s}_{11}^*\Psi}{\tilde{s}_{13}^*}=\left\{\begin{array}{lr}\tilde{s}_{13}\psi_1^*,& R\rightarrow 0,\\
\psi_3-(\tilde{s}_{13}\tilde{s}_{11}^*/\tilde{s}_{13}^*)\psi_3^*,&R\rightarrow \infty,\end{array}\right.
\end{equation}
the right hand side of which defines the second ($i=3$) row of the matrix $\tilde{s}_{ij}$. Namely, $\tilde{s}_{31}=\tilde{s}_{13}$,  $\tilde{s}_{33}=-(\tilde{s}_{13}\tilde{s}_{11}^*/\tilde{s}_{13}^*)$, and $\tilde{s}_{3j} = 0$ for $j>3$.

\begin{figure}
\centerline{\includegraphics[width=\columnwidth]{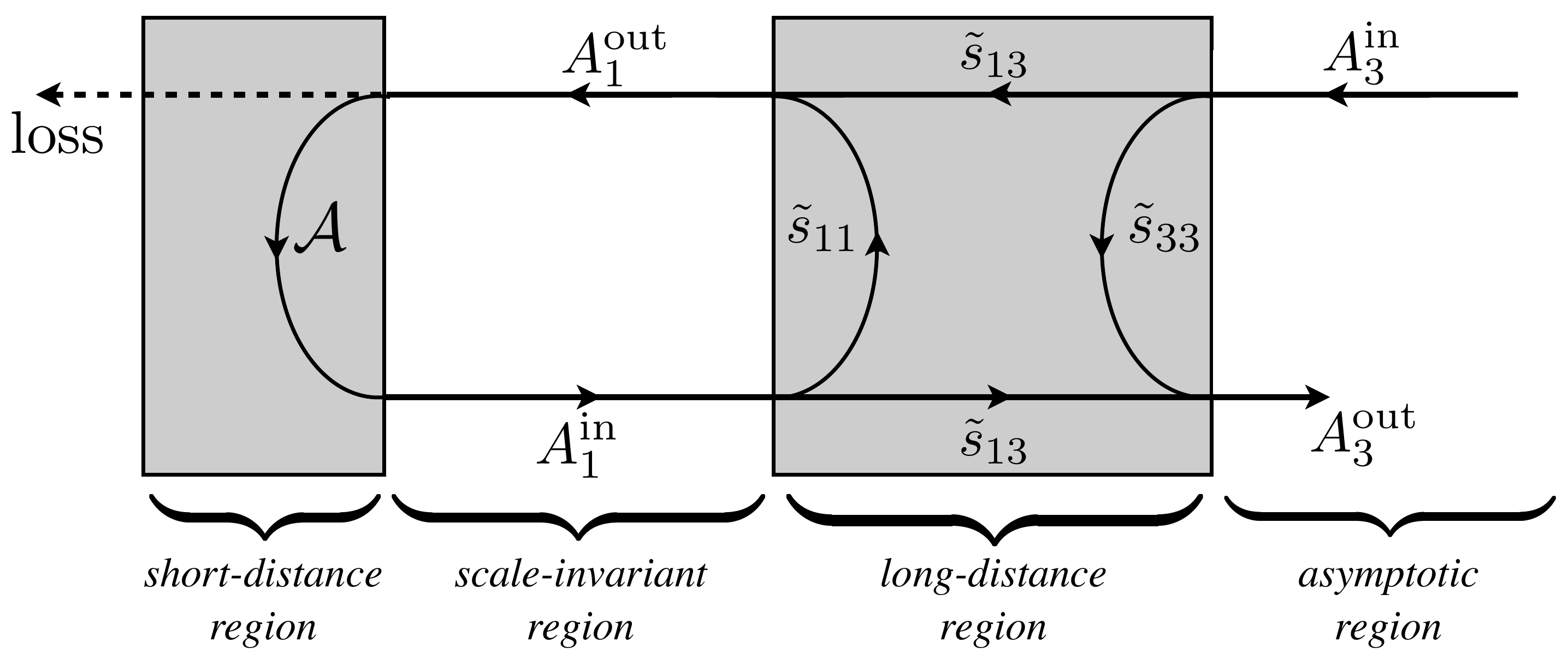}}
\caption{A three-body wave arriving from large hyperradius $R$ with amplitude $A^{\rm in}_3$ in the triatomic channel $i=3$ can follow various pathways before it either returns to large $R$, or gets lost at $R\sim R_e$ by turning into an atom and a deep dimer. One can imagine a Fabry-Perot interferometer, the mirrors of which are formed by the {\it short-distance} and {\it long-distance} regions. Multiple reflections by these regions can lead to the resonant denominator in the three-body loss rate formula.} 
\label{fig:fabry}
\end{figure}

Because $\tilde{s}$ is unitary one sees that $\tilde{s}_{i1}=\tilde{s}_{i3}=0$ for $i>3$. Therefore, the upper left $2\times 2$ block completely separates from the rest of the matrix. The problem of calculating the three-body loss rate then reduces to the problem of finding the four amplitudes $A_1^{\rm in/out}$ and $A_3^{\rm in/out}$, which are the coefficients in front of the corresponding incoming and outgoing waves in the three-body wavefunction, see Fig.~\ref{fig:fabry}. The coefficient $A_3^{\rm in}$ is found by projecting the initial correctly normalized six-dimensional plane wave into the state $\psi_3$. The amplitudes $A_1^{\rm in/out}$ are related by the three-body contact condition $A^{\rm in}_1 = \mathcal{A}\,A^{\rm out}_1$ where $\mathcal{A}= -(kR_0)^{-2is_0}e^{-2\eta_*}$. Finally, the relation between the incoming and outgoing amplitudes given by the matrix $\tilde{s}_{ij}$ provides the last two linear equations necessary to solve the problem: $A_1^{\rm out} = \tilde{s}_{11} A_1^{\rm in}+\tilde{s}_{13} A_3^{\rm in}$ and $A_3^{\rm out} = \tilde{s}_{31} A_1^{\rm in}+\tilde{s}_{33} A_3^{\rm in}$. The loss rate is then obtained by calculating the difference between incoming and outgoing fluxes either for $R\rightarrow \infty$ or $R\rightarrow 0$. Averaging over the thermal distribution one recovers the formula for $L_3$ presented in Eq.~(4) of the main text, where we now see that the non-trivial $k$-dependence of the integrand comes from the interference between the various pathways represented in Fig.~\ref{fig:fabry}. The simplified approximate formula given in Eq.~(5) of the main text corresponds to neglecting any reflection from the long-distance region, hence no more interferences and no log-periodic modulation of $L_3 T^2$ with $\lambda_{\rm th}/R_t$.

\section{Matrix $s_{ij}$ at unitarity}

At unitarity, the procedure of transforming the matrix $s_{ij}$ into block diagonal form is very simple. Having an infinite $a$ does not introduce a lengthscale into the problem and, as a consequence, the adiabatic hyperangular eigenfunctions do not depend on the hyperradius, leading to the complete separability \cite{Efimov} between the hyperangular and hyperradial problems. Namely, the three-body wavefunction can be written as
\begin{equation}\label{EfimovDecomposition}
\Psi({\bf R}) = \sum_s \phi_s(\hat{R}) F_s(R)R^{-2}, 
\end{equation}
where $\phi_s(\hat{R})$ and $s^2$ are, respectively, the (normalized) eigenfunctions and eigenvalues of the hyperangular kinetic energy operator supplemented with the unitary two-body contact conditions. The hyperradial wavefunctions satisfy
\begin{equation}\label{eq:schro_R}
\left(-\frac{d^2}{d R^2} - \frac{1}{R} \frac{ d}{d R} +
\frac{s^2}{R^2} \right) F_s(R) =k^2\, F_s(R).
\end{equation}
In the case of three identical bosons considered here, the set $\{s\}$ contains a single imaginary number $s = i\,s_0 \simeq i 1.00624$ (Efimovian sector) and an infinite number of real numbers (non-Efimovian sectors). In the Efimovian sector the attractive $-s_0^2/R^2$ potential gives rise to the following asymptotic behavior of $F_{is_0}$: for $R\ll 1/k$ we have $F_{is_0}(R)\propto R^{\pm is_0}$ and in the opposite limit 
$F_{is_0}(R)\propto \exp(\pm ikR)/\sqrt{R}$. These two asymptotes of the same function actually define the two channels $i=1$ and $i=3$ discussed in Sec. \ref{sec1}, the hyperangular wavefunctions being $\Phi_1=\tilde{\Phi}_3=\phi_{is_0}$. The rest of $\phi_s$, appropriately relabelled, form the rest of the set $\tilde{\Phi}_i$. The corresponding matrix $\tilde{s}_{ij}$ has a $2\times 2$ block in its upper left corner, which describes the transmission and reflection of the wavefunction $F_{is_0}(R)$ by the {\it long-distance} region $R\sim 1/k$. The rest of $\tilde{s}_{ij}$ is simply diagonal because (i) these channels are decoupled from each other and (ii) the repulsive $s^2/R^2$ potentials do not allow (in the zero-range approximation) for a transmission of the corresponding waves to the {\it short-distance} region $R\sim R_e$. Solutions of Eq.~(\ref{eq:schro_R}) can be written in terms of Bessel functions. In particular, the wave that has properties of Eq.~(\ref{Psi}) can be written by setting
\begin{equation}\label{Fis0}
F_{is_0}(R)=\frac{2^{is_0}\Gamma(1+is_0)}{\sqrt{2s_0}}[J_{is_0}(kR)-e^{-\pi s_0}J_{-is_0}(kR)]
\end{equation}
in Eq.~(\ref{EfimovDecomposition}). Expanding Eq.~(\ref{Fis0}) at small $R$ we obtain the result for $s_{11}(\infty)$, which has already been mentioned. For completeness, from the large-$R$ asymptotes of $J_{\pm is_0}$ we get $\tilde{s}_{13}=2^{is_0}\sqrt{2/\pi s_0}\,\Gamma(1+is_0)\sinh(\pi s_0)\exp(-\pi s_0/2-i\pi/4)$. 

\section{Matrix $s_{ij}$: positive $a$}

In the case $a>0$ we have to take into account another discrete channel: the large-$R$ atom-dimer one denoted by $i=2$. By using a similar construction as in the case of negative $a$, one can show that the matrix $s_{ij}$ can be reduced to a block-diagonal form with a $4\times 4$ block in the upper left corner, i.e., there are actually two triatomic channels, $i=3$ and $i=4$, coupled to the atom-dimer and Efimov ones, and decoupled from the rest of the triatomic continuum, $i> 4$. As we have shown in Sec.~\ref{sec2}, $s_{12}(ka)\rightarrow 0$ as $a\rightarrow\infty$. It is then straightforward to show that in this limit channels 1 and 3 approximately decouple from channels 2 and 4. This means that dimers existing in the system are more likely to break-up or scatter elastically \cite{remarks22} than to relax to deeply bound states. This is consistent with our earlier conclusion on the atom-dimer chemical quasi-equilibrium close to the resonance.

\section{Evaporation and anti-evaporation}

For $a\,{<}\,0$, the three-body recombination to deeply bound states gives the contribution $\dot{N}_{3body}=-\int L_3 n^3({\bf r})d^3 r$ to the atomic decay, and the corresponding energy loss rate equals
\begin{widetext}
\begin{equation}\label{Edot3body}
\dot{E}_{3body}=-\int d^3 r\,\left\{\frac{L_3 n^3({\bf r})}{3}\left[3U({\bf r})+\frac{3k_B T}{2}\right]+\frac{n^3({\bf r})}{3}\frac{72\sqrt{3}\pi^2\hbar(1-e^{-4\eta_*})}{mk_T^6}\int_0^\infty\frac{\hbar^2k^2}{m}\frac{(1-|s_{11}|^2)e^{-k^2/k_T^2}kdk}{|1+(kR_0)^{-2is_0}e^{-2\eta_*}s_{11}|^2}\right\}
\end{equation}
\end{widetext}
where $L_3 n^3({\bf r})/3$ is the frequency of three-body events per unit volume, $3U({\bf r})$ and $3k_B T/2$ are the loss of trapping potential energy and of center-of-mass kinetic energy by each recombining triple, and the last term is the loss of relative-motion kinetic energy~\cite{note3}. Let us write the lost energy per lost atom as $\dot{E}_{3body}/\dot{N}_{3body}~{=}~(3~{-}~\delta) k_B T$ where $\delta \, k_B T$ is the excess energy as compared to the average energy per atom $3 k_B T$. 

For evaporation, $\dot{E}_{evap}/\dot{N}_{evap}\approx (\eta+\kappa)k_B T$, where we can take the expression of $\kappa$ in terms of $\eta$ for a harmonic trap and given in terms of incomplete gamma functions in~\cite{luiten1996}. Indeed, as realized in~\cite{luo2006evap}, two-body collisions leading to an evaporative loss occur mainly in the cloud center where the trap is harmonic, and the relative momentum for such a collision is approximately fixed by the trap depth so that the result derived in~\cite{luiten1996} for an energy-independent two-body cross-section is applicable. Typically, we have $\kappa\,{\simeq}\,0.68$ for $\eta\,{=}\,6$ and $\kappa\,{\simeq}\,0.78$ for $\eta\,{=}\,8$.
%For evaporation, $\dot{E}_{evap}/\dot{N}_{evap}\approx (\eta+\kappa)k_B T$ with $\kappa=(\eta-5)/(\eta-4)$~\cite{luo2006evap,note2}.
The condition of constant temperature means that $\dot{E}_{evap}+\dot{E}_{3body}=3k_B\,T\,(\dot{N}_{evap}+\dot{N}_{3body})$,
which yields $\dot{N}_{evap}/\dot{N}_{3body}=\delta/(\eta+\kappa-3)$. At unitarity, we can neglect $s_{11}$ in~(\ref{Edot3body}), which gives $\delta\approx5/3$. For $-a\,{\ll}\,\lambda_{\rm th}$, we recover $\delta\,{=}\,1$ as in \cite{weber2003three}.

For $a\,{>}\,0$, we use $\delta\,{=}\,5/3$ when $a\,{>}\,\lambda_{\rm th}$. In the opposite limit $a\,{\ll}\,\lambda_{\rm th}$ we use $\delta\,{=}\,1$ from \cite{weber2003three}.

\section{Discussion of Uncertainties}

We make use of the grand-canonical equation of state for a degenerate Bose gas in the mean-field limit to calibrate our measurement of the value of $\lambda_{3}$~\cite{Navon2011}.  We produce a condensate at $a=200\,a_{0}$ and measure the normalized pressure $h$ versus the gas parameter $\nu= \frac{\mu}{g}a^{3}$, where $g=4\pi \hbar^2a/m$. Next, we find that in order to match $h(\nu)$ to the mean-field prediction, we must multiply the pressure by a constant $\xi=2.45$.  $\xi$ corrects for errors in the calibrations of our experimental system, \emph{e.g.} our absolute atom counting, through the product ${\omega_{r}}^{2}/({\omega_{z}}^{4}\,{(px)}^{3}\,\sigma_{\rm A})$, where $px$ is the size of a camera pixel magnified through the imaging system to the gas location and $\sigma_{\rm A}$ is the atomic absorption cross-section for imaging light of finite linewidth. When fitting Eq.~(3) to our data, we extract two fit parameters: $\gamma_{3}=A(T)L_{3}(T)N^{2}(0)$ and $N(0)$. Consequently, our result for $\lambda_{3}=L_{3}T^{2}\propto \gamma_{3}T^{5}/({N^{2}(0)}{\omega_{r}}^{4}{\omega_{z}}^{2})$ scales as ${\omega_{z}}^{8}\,{(px)}^{6}\,{\sigma_{\rm A}}^{2}/{\omega_{\rho}}^4$.  This factor is exactly $\xi^{-2}$.  Ultimately, we estimate our uncertainty in $\xi^{-2}$ to be $25\,\%$, dominated by the uncertainty in our trap frequency and pixel size.  For the data in Fig.~3, we have an additional $20\,\%$ uncertainty arising from the $T^{2}$ scaling with $(px)^{4}$.